\documentclass[10pt]{revtex4}
 
 \usepackage{hyperref}
 \usepackage{amsmath,amsfonts,amssymb}
 \hypersetup{dvips,dvipdfm,colorlinks=true,urlcolor=magenta,filecolor=magenta,linktoc=page,citecolor=red,linkcolor=blue,bookmarks=true}
 \usepackage{graphicx,epstopdf}

\usepackage{color}

\begin{document}
\title{Cosmological consequences in the framework of Generalized Rastall theory of gravity }

\author{Dipanjana Das$^1$\footnote {ddipanjana91@gmail.com}}
%\author{Supriya Pan$^2$\footnote {span@iiserkol.ac.in}}
\author{Sourav Dutta$^2$\footnote {sduttaju@gmail.com}}
\author{Subenoy Chakraborty$^1$\footnote {schakraborty.math@gmail.com}}
\affiliation{$^1$Department of Mathematics, Jadavpur University, Kolkata-700032, West Bengal, India\\
		%$^2$ %Department of Mathematics, Raiganj Surendranath Mahavidyalaya,
		%Sudarshanpur, Raiganj, West Bengal 733134, India\\
	$^2$ Department of Pure Mathematics, University of Calcutta, 35, Ballygunge Circular Rd, Ballygunge, Kolkata, West Bengal 700019}

\begin{abstract}
The paper deals with generalized Rastall theory of gravity and its cosmological consequences in the background of homogeneous and isotropic flat FLRW model with perfect fluid as the matter context. The model shows a non singular era (emergent scenario) at the early phase of expansion for a particular choice of the Rastall parameter. Also the model finds to be equivalent to the particle creation mechanism in Einstein gravity in the framework of non-equilibrium thermodynamics. Universal thermodynamics is briefly presented and it is found that the entropy function in Rastall theory is the usual Bekenstein entropy and there is no correction to it. Finally, a complete cosmic history starting from inflation to late time acceleration is presented for suitable choices of the Rastall parameter.\\

\end{abstract}

%----------------------------------------------------------------------------------------
\maketitle
%%%%%%%%%%%%%%%%%%%%%%%%%%%%%%%%%%%%%%%%%%%%%%%%%%%%%%%%%%%%%%%%%%%%%%%%%%%%%%%%%%%%%%%%%%%%%%%%%%%%%%%%%%%%%%%%%%%%%%%%%%%%
\section{Introduction}
The covariant conservation of the energy momentum tensor is a well known ingradient of Einstein's theory of gravity. Further, due to Noether's theorem this conservation of matter leads to conservation of globally defined quantities which are expressed as integrals of the energy-momentum tensor over some space-like hypersurface. Hence in general relativity, the total rest energy  (mass) of a system remains conserved, but there is no experimental evidence in favour of this assertion. As a result, there are proposals for modification of general relativity which do not care about this covariant conservation of energy-momentum tensor. Long back in 1972 Rastall \cite{Rastall:1976uh,Rastall:1973} proposed such a modified theory of gravity and recently it has been discussed much in the literature. In this theory, the matter source is described by the energy momentum tensor as in general relativity and also by the metric of the external space. So in empty space this modified gravity theory coincides with Einstein gravity and hence it agrees with Mach's principle \cite{Majernik:2006jg}.\\

  The phenomenological nature of Rastall theory and absence of an action from which Rastall theory can be derived, are the major drawbacks of this modified theory while the rich structure of it can be associated with many fundamental aspects of a gravity theory. In curved space--time due to quantum effects, the classical energy momentum tensor is modified by quantities related to the curvature of the space-time \cite{Birrell:1982}. Also in space-times bounded by horizon due to propagation of quantum fields there is violation of classical conservation law (Chirality of the quantum modes) which is termed as gravitational anomaly \cite{Bertlmann:2000}. Hence, one may consider Rastall theory as a phenomenological way of determining the effects of quantum fields in curved space-time in a covariant way. Further, although there is no action from which Rastall theory can be derived but it is possible to have an action if an external field is introduced in the Einstein-Hilbert action through some Lagrange multiplier. Moreover, geometrical frameworks like Weyl geometry may lead to similar equations as Rastall theory \cite{Nuovo:1984, Almeida:2013dba}.\\
  
  The Rastall gravity theory has been extensively used in various cosmological aspects \cite{Batista:2012hv, Fabris:2012hw, Batista:2011nu}. Different cosmological scenarios are studied in the present work for this modified gravity theory mostly in the framework of particle creation mechanism. The particle creation process \cite{Gibbons:1977mu,Parker:1971pt,Ford:1986sy, Batista:2011nu,Pereira:2009kv} and this modified gravity theory have similarity in cosmic evolution as both of them do not respect the conservation of the energy momentum tensor. In fact, in the present work we consider the generalized Rastall gravity with variable Rastall parameter. Emergent scenario is obtained for a particular choice of the Rastall parameter. Then it is found that a complete cosmic scenario is possible starting from inflation to the present late time acceleration by suitable choices of the Rastall parameter. Also Universal thermodynamics in Rastall gravity is briefly presented. The paper is organized as follows: 
  Basic equations are presented in section II, section III deals with singular and non singular cosmological solutions. The equivalence of Rastall gravity with particle creation formulation in Einstein Gravity are shown to be equivalent in section IV. Also a continuous cosmic evolution has been presented in this section. Section V deals with universal thermodynamics in this modified gravity theory. The paper ends with conclusion in section VI.

\section {Basics Equation for Rastall Gravity}

In most of the gravity theories, the source of energy-momentum is described by a divergence free tensor which couples to the geometry in a minimal way. Although the resulting energy-momentum tensor satisfies the conservation law but it is not obeyed by the particle production process. So it is desirable to consider a known divergence free energy-momentum tensor in the modified gravity theory. Rastall theory is an example of such a modification to the gravity theory. In this theory the usual energy-momentum conservation law is not satisfied rather, it satisfies the following relation
\begin{equation}
T^{\mu \nu}_{;\mu}=lR_{,\mu}, \label{r1}
\end{equation}
where $R$ and $l$ are respectively the usual Ricci's scalar and a constant known as Rastall constant. One may interpret the constant parameter $l$ as a measure of the tendency of the geometry to couple with matter fields or vice-versa. Hence the Einstein's field equation can be written as -
\begin{equation}
G_{\mu \nu}+\xi_{0}lg_{\mu\nu}R=\xi_{0}T_{\mu\nu},\label{r2}
\end{equation}
i.e.,

\begin{equation}
G_{\mu \nu}=\xi_{0}S_{\mu\nu},\label{r3}
\end{equation}
where $\xi_{0}$ is the gravitational coupling constant to the Rastall theory and $S_{\mu \nu}$ is termed as effective energy-momentum tensor having expression
\begin{equation}
S_{\mu\nu}=T_{\mu \nu}-\frac{\xi_{0}lT}{4\xi_{0}l-1}g_{\mu \nu},\label{r4} 
\end{equation}

Note that here the matter fields and the geometry are non-minimally coupled and it is compatible with some observational data \cite{Riess:1998}, \cite{sp}. As the usual particle production process during cosmic evolution does not satisfy the energy-momentum conservation law so, one may consider Rastall theory as classical background formulation of this phenomena.\\

Taking trace of equation (\ref{r2}), we have:
\begin{equation}
R\frac{(4\xi_{0}l-1)}{T}=\xi_{0},\label{r5} 
\end{equation}
where $T$ is the trace of the energy-momentum tensor. From the above relation, it is clear that $\xi_{0}l\neq\frac{1}{4}$ \cite{Moradpour:2017}, \cite{Rastall:1973}. Subsequently this theory has been modified (known as Generalized Rastall Theory) in which the Rastall parameter $l$ is no longer a constant. So, the divergence of the energy-momentum tensor takes the form
\begin{equation}
T^{\mu \nu}_{;\mu}=(\lambda R)^{,\nu},\label{r6} 
\end{equation}
where $\lambda$ is termed as varying Rastall parameter.
As a consequence due to Bianchi Identity $(G^{;\nu}_{\mu \nu}=0)$ the Einstein's equation now takes the form-
\begin{equation}
G_{\mu\nu}+\xi\lambda g_{\mu \nu}=\xi T_{\mu \nu}.\label{r7} 
\end{equation} with $\xi$, a constant.

One may note that although the Einstein's field equations (\ref{r2}) and (\ref{r7}) apparently look identical but here $\lambda$ is not a constant, measures the coupling between geometry and matter field. However, the original Einstein's field equations can be obtained in the limit $\lambda\rightarrow 0$ (The limit implies the matter field and geometry are couple to each other in a minimal way.) Also in the above modified Einstein's field equations the Rastall parameter $\lambda$ can be interpreted as a varying cosmological constant in Einstein Gravity.\\\\

In the present work we consider homogeneous and isotropic flat FLRW space time model having line element 
\begin{equation}
ds^2=-dt^2+a^2(t)\big[dr^2+r^2 \big(d\theta^2+\sin^2 \theta d\phi^2 \big)\big],\label{r8} 
\end{equation} 
where $a(t)$ is the usual scale factor and the curvature scalar $k$ is assumed to be $0$ for flat model. Also if it is assumed that the matter source in the Universe is a perfect fluid with barotropic equation of state, then the modified Friedmann equations in  generalized Rastall theory are given by 
\begin{equation}
(12\xi \lambda -3)H^2+6\xi \lambda H=-\xi \rho,\label{r9} 
\end{equation}
\begin{equation}
(12\xi \lambda-3)H^2+(6\xi \lambda-2)\dot H=\xi p,\label{r10} 
\end{equation}
where an over dot indicates derivative with respect to the cosmic time $t$. Also the explicit form of the energy conservation equation(\ref{r6}) can be written as

\begin{equation}
\frac{d}{dt}\left(\frac{\rho+6\lambda\dot H}{1-4\lambda \xi}\right)+3H(\rho+p)=0.\label{r11} 
\end{equation}

It is to be noted that equations (\ref{r9})--(\ref{r11}) are not independent, infact equation (\ref{r11}) can be obtained from equations (\ref{r9}) and (\ref{r10}). If we now rewrite the above modified Friedmann equations (\ref{r9})and(\ref{r10}) as the usual Friedmann equations, then we have\\\\
\begin{equation}
3H^2=\xi (\rho+\rho_e),\label{r12} 
\end{equation}
and
\begin{equation}
2\dot H=-\xi(\rho+\rho_{e}+p+p_{e}),\label{r13} 
\end{equation}
where $\rho_{e}=12\lambda H^2+6\lambda \dot H$ and $ p_{e}=-(12\lambda H^2+6\lambda \dot H)$ are the additional energy density and thermodynamic pressure due to generalized Rastall theory. As $ p_{e}=-\rho_{e}$ so the additional matter field in the Rastall theory behave as varying cosmological parameter in Einstein gravity. In other words one may consider that generalized Rastall theory is equivalent to Einstein gravity with a varying cosmological parameter. This supports the claim in \cite{Visser:2017gpz} even for generalized Rastall gravity.

\section{singular and Non-Singular Universe in Rastall gravity}

In this section it will be examined whether it is possible to have a non-singular cosmological solution (i.e., emergent scenario) for the above generalized Rastall theory. From the field equations (\ref{r9}) and (\ref{r10}) one obtains (after some simplifications):
\begin{equation}
\frac{2\dot H}{3 H^2}=\gamma \frac{(1-4\xi \lambda )}{(3\xi\lambda \gamma -1)},\label{r14} 
\end{equation}
where $w=\gamma-1$ is the equation of state parameter of the perfect fluid. Now choosing the Rastall parameter $\lambda$ as 
\begin{equation}
\lambda=\frac{1+d_0 H}{3\xi \gamma},\label{r15} 
\end{equation} 
($d_0$, an arbitrary constant), the cosmic evolution is given by
\begin{equation}
2\dot H=\frac{(3 \gamma -4)}{d_0} H-4H^2.\label{r16}       
\end{equation}
Now, depending on the sign of $\beta=\frac{(3 \gamma -4)}{d_0}$, the possible solutions are\\

{\bf Case-I:}~$\beta>0$,~i.e.,~$\gamma>\frac{4}{3}$
\begin{eqnarray}  
\big(\frac{a}{a_{0}}\big)^2&=&\big[1+\frac{4H_0}{\beta}\{e^{\frac{\beta}{2}}(t-t_0)-1\}\big], \nonumber\\
\frac{H}{H_{0}}&=&\frac{\frac{\beta}{4}}{\big[H_0-(H_0-\frac{\beta}{4})e^{-\frac{\beta}{2}(t-t_0)}\big]}. \label{r17}
\end{eqnarray}
\begin{center}
	or
\end{center}
\begin{eqnarray}  
\big(\frac{a}{a_{0}}\big)^2&=&1+e^{4H_0(t-t_0)}, \nonumber\\
\frac{H}{H_{0}}&=&\frac{2}{\big[1+e^{-4H_0(t-t_0)}\big]}. \label{r18}
\end{eqnarray}
according as $H>\frac{\beta}{4}$ or $H<\frac{\beta}{4}$.\\

{\bf Case-II:}~$\beta=0$,~i.e.,~$\gamma=\frac{4}{3}$

\begin{eqnarray}  
\big(\frac{a}{a_{0}}\big)^2&=&1+2H_0(t-t_0), \nonumber\\
\frac{H_0}{H}&=&1+2H_0(t-t_0). \label{r18(i)}
\end{eqnarray}

{\bf Case-III:}~$\beta<0$,~i.e.,~$\gamma<\frac{4}{3}$
\begin{eqnarray}  
\big(\frac{a}{a_{0}}\big)^2&=&\big(1+\frac{4H_0}{\mu^2}\big)-\frac{4H_0}{\mu^2}e^{-\frac{\mu^2}{
		2}(t-t_0)}, \nonumber\\
\frac{H}{H_0}&=&\frac{\frac{\mu^2}{
		4}}{\big[\big(H_0+\frac{\mu^2}{
		4}\big)e^{\frac{\mu^2}{
			2}(t-t_0)}-H_0\big]}. \label{r18(ii)}
\end{eqnarray}
with $\mu^2=-\beta$.\\

In the above solutions we have $t_0$ and $a_0$ are integration constants with $a=a_0$, $H=H_0$ at $t=t_0$. Among the above solutions, the solution (\ref{r18}) represents a non-singular model of the Universe. For this  emergent solution we have
$$(i) a\rightarrow a_0,~H\rightarrow 0 ~\mbox{as}~ t\rightarrow -\infty$$
$$(ii) a\sim a_0,~H\sim 0 ~\mbox{for}~ t<< t_0$$
$$(iii) a\sim \exp\{4H_0(t-t_0)\},~H\sim 2H_0 ~\mbox{for}~ t>> t_0.$$

Thus it is possible to have a scenario of emergent Universe in generalized Rastall theory. The other three solutions represented by equations (\ref{r17}), (\ref{r18(i)}) and (\ref{r18(ii)}) represent big-bang model of the Universe. The big-bang singularity occurs at 
\begin{equation}
t_c=\left\{
\begin{array}{lll}
t_0+\frac{2}{\beta}\ln |1-\frac{\beta}{4H_0}| & \mbox{ for the solution (\ref{r17}) } \\
t_0-\frac{1}{2H_0} & \mbox{ for the solution (\ref{r18(i)}) } \\
t_0-\frac{2}{\beta}\ln |\frac{H_0}{H_0-\frac{\beta}{4}}| & \mbox{ for the solution (\ref{r18(ii)}) }\\
\end{array}
\right\},\nonumber
\end{equation}

The Universe evolves undoubtedly in an exponential manner after the big-bang for the solution (\ref{r17}) and in a power law form for the solution (\ref{r18(i)}) while there is finite evolution for the solution (\ref{r18(ii)}).

\section{Particle Creation in Rastall Theory: continuous Cosmic Evolution}
This section is an attempt to show whether the generalized  Rastall theory can be considered as a particle creation formulation in Einstein Gravity and it is examined whether a continuous cosmic evolution is possible with proper choices of the Rastall parameter. From the usual Friedmann equations with particle creation mechanism the cosmic evolution can be described as \cite{Chakraborty:2014a}
\begin{equation}
\frac{2\dot H}{3 H^2}=\gamma(\frac{\Gamma}{3H}-1),\label{r19}
\end{equation}
where, $\Gamma$ is the particle creation rate of the cosmic fluid particles. Thus, comparing (\ref{r19}) with the evolution equation(\ref{r14}) in generalized Rastall Theory, We have-
\begin{equation}
\Gamma=\frac{3 H\xi \lambda(3\gamma-4)}{(3\xi\lambda\gamma-1)}.\label{r20}
\end{equation}

Hence, Generalized Rastall Theory may be considered as a particle creation mechanism in Einstein Gravity and the particle creation rate is determined by the Rastall parameter $\lambda$ in equation (\ref{r20}). Thus generalized Rastall theory may be considered as Einstein gravity with imperfect fluid (i.e., viscous).\\
%\section{continous Cosmic Evolution in Rastall Theory}
Due to the variable nature of the parameter $\lambda$ in the generalized Rastall theory, it will be examined whether for a continuous choice of $\lambda$, it is possible to have a complete cosmic evolution from early inflationary era to present late time acceleration. In previous section it is seen that with proper choice of $\lambda$ (phenomenological) it is possible to have cosmological solution for emergent scenario.\\

In an earlier work \cite{Chakraborty:2014a}, \cite{Pan:2014} it has been shown that it is possible to choose the particle creation rate from thermodynamical point of view, so that the entire cosmic evolution can be obtained in FLRW model. Inspired by that work in the following there are choices for the parameter $\lambda$ for which three phases of cosmic evolution namely early inflationary era, intermediate matter dominated era and finally late time acceleration phase can be obtained. Here it should be noted that similar to particle creation mechanism there is no physical ground for these choices for $\lambda$ , they are purely phenomenological in nature.\\

Let $t_1$ be the time instant in which Universe evolves from inflationary era to the matter dominated era. Similarly $t_2>t_1$ be the time at which the Universe makes a transition to the late time acceleration. Let $(a_1, H_1)$ and $(a_2, H_2)$ be the scale factor and Hubble parameter at these time instants.\\

{\bf Early Inflationary Era: ($t<t_1$)}\\

The choice for $\lambda$ and the cosmic solution is the following
$$\lambda=\frac{\eta \big(\frac{H}{H_1}\big)}{\xi\big[3\eta \gamma\frac{H}{H_1}-(3\gamma -4) \big]},$$

$$\big(\frac{a}{a_1}\big)^{\eta} e^{2\frac{(1-\eta)}{3\gamma}\big[\big(\frac{a}{a_1}\big)^{\frac{3\gamma}{2}}-1\big]}=e^{H_1(t-t_1)},$$
$$\frac{H}{H_1}=\big[\eta\{Lambert~ W\big(\frac{1-\eta}{\eta}\exp\{\frac{2(1-\eta)+3\gamma H_1(t-t_1)}{2\eta} \}\big) +1 \}\big]^{-1},$$
$$\frac{\rho}{\rho_1}=\big[\eta\{Lambert~ W\big(\frac{1-\eta}{\eta}\exp\{\frac{2(1-\eta)+3\gamma H_1(t-t_1)}{2\eta} \}\big) +1 \}\big]^{-2},$$
%$$\frac{\eta}{\eta_1}=\big[\eta\{Lambert~ W\big(\frac{1-\beta}{\beta}\exp\{\frac{2(1-\beta)+3\gamma H_1(t-t_1)}{2\eta} \}\big) +1 \}\big]^{-\frac{2}{\gamma}},$$
%$$\frac{T}{T_1}=\big[\eta\{Lambert~ W\big(\frac{1-\beta}{\beta}\exp\{\frac{2(1-\beta)+3\gamma H_1(t-t_1)}{2\eta} \}\big) +1 \}\big]^{-\frac{2(\gamma-1)}{\gamma}}.$$

where $\eta$ is a constant and the function $Lambert~W(x)$ is defined by $Lambert~W(x) e^{Lambert~W(x)}=x$.\\

{\bf Intermediate matter Dominated Era: ($t_1<t<t_2$)}\\

Here we choose $\lambda=\lambda_0$, a constant, where $\lambda_0=\frac{\frac{\eta}{\xi}}{3\eta \gamma-(3\gamma-4)}$ and the cosmological solutions for this phase is given by\\
$$\frac{a}{a_1}=\big[1+\frac{3\gamma}{2}(1-\eta)H_1(t-t_1)\big]^{\frac{2}{3\gamma(1-\eta)}},$$
$$\frac{H}{H_1}=\big[1+\frac{3\gamma}{2}(1-\eta)H_1(t-t_1)\big]^{-1},$$
$$\frac{\rho}{\rho_1}=\big[1+\frac{3\gamma}{2}(1-\eta)H_1(t-t_1)\big]^{2},$$
%$$\frac{\eta}{\eta_1}=\big[1+\frac{3\gamma}{2}(1-\eta)H_1(t-t_1)\big]^{\frac{2}{\gamma}},$$
%$$\frac{T}{T_1}=\big[1+\frac{3\gamma}{2}(1-\eta)H_1(t-t_1)\big]^{\frac{2(\gamma-1)}{\gamma}}.$$

{\bf Late Time Acceleration: ($t>t_2$)}\\
If we choose $\lambda=\frac{\big[\frac{\psi H_2}{\xi}\big]}{\big[3\gamma \psi H_2-(3\gamma-4)H^2\big]}$, then the cosmological parameters are given by\\
$$\big(\frac{a}{a_2}\big)^{\frac{3\gamma}{2}}=\frac{1}{\sqrt{\psi H_2}}\sinh \big[\sqrt{\psi H_2}(t-t_0)\big], $$
$$H=\sqrt{\psi H_2}\coth \big[\sqrt{\psi H_2}(t-t_0)\big], $$
$$\frac{\rho}{\rho_2}=\frac{\psi}{H_2}\coth^2\big[\sqrt{\psi H_2}(t-t_0)\big], $$
%$$\frac{\eta}{\eta_2}=\big[\frac{\psi}{H_2}\coth^2\big[\sqrt{\psi H_2}(t-t_0)\big]\big]^{\frac{1}{\gamma}},$$
%$$\frac{T}{T_2}=\big[\frac{\psi}{H_2}\coth^2\big[\sqrt{\psi H_2}(t-t_0)\big]\big]^{\frac{\gamma-1}{\gamma}}.$$

From the above solutions one can easily verify the continuity of the parameter $\lambda$  and all the geometrical and physical parameters presented above across $t=t_1$ while there will be continuity at $t=t_{2}$, provided the arbitrary parameters involved in the solutions are constrained by the following relations.\\

%For continuity at $t=t_2:$

$$\lambda:~\psi=\eta H_2$$
$$H:~\big(\frac{a_1}{a_2}\big)^{\frac{3\gamma(1-\eta)}{2}}=\frac{H_2}{H_1},~H_2^2=\psi H_2+1$$
$$~\mbox{\i.e.,}~H_2^2(1-\eta)=1.$$

%%%%%%%%%%%%%%%%%%%%%%%%%%%%%%%%%%%%%%%%%%%%%%%
\begin{figure}
	\centering
	\includegraphics[width=0.5\textwidth]{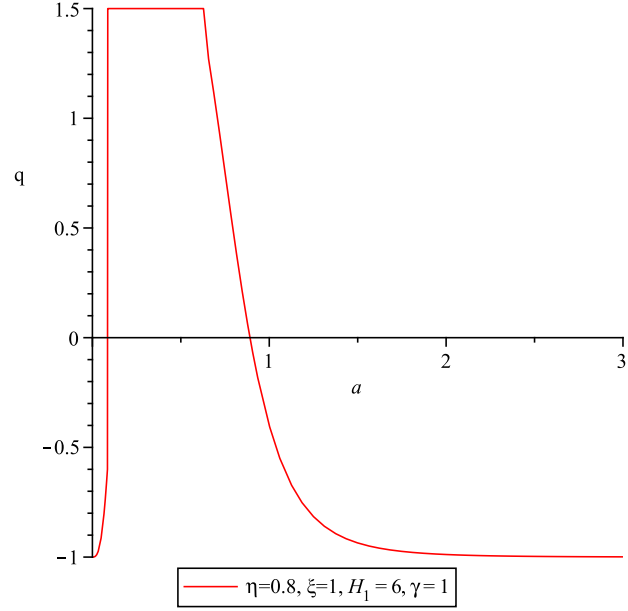}\\
	\caption{Shows the evolution of the deceleration parameter against the scale factor $a$}
	\label{fig1}
\end{figure}
%%%%%%%%%%%%%%%%%%%%%%%%%%%%%%%%%%%
\begin{figure}
	\centering
	\includegraphics[width=0.5\textwidth]{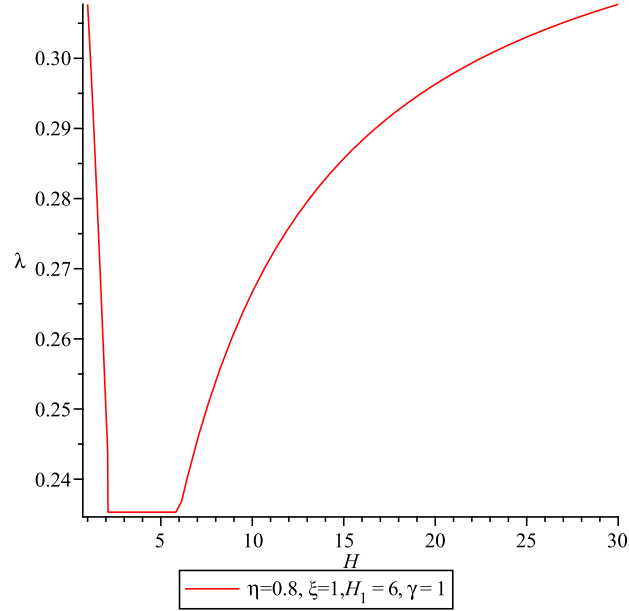}\\
	\caption{Continuous evolution of the Rastall parameter $\lambda$}
	\label{fig2}
\end{figure}

%%%%%%%%%%%%%%%%%%%%%%%%%%%%%%%%

\section{Universal Thermodynamics in Rastall Theory of Gravity}

In this section, we consider universal Thermodynamics with Universe as a non--stationary gravitational system. As from cosmological point of view the homogeneous and isotropic  FLRW Universe may be considered as dynamical spherically symmetric spacetime so there is only inner trapping horizon which for the present model coincides with the apparent horizon. Using Hayward's \cite{Hayward:1993,Hayward:1994,Hayward:1998,Hayward:2004,Hayward:1997} thermodynamical analysis for dynamical BH one may consider unified first law to study universal thermodynamics. In this context, Cai and Kim \cite{Cai:2005} derived the usual Friedmann equations from the Clausius relation $\delta Q=T ds$ at the apparent horizon of the FLRW Universe, considering Hawking temperature and Bekenstein entropy on the (apparent) horizon as 

\begin{equation}
T_H=\frac{1}{2 \pi R_A},~~S_B=\frac{\pi R_A^2}{G},\label{t1} 
\end{equation}
with $R_A$ is the geometric radius of the apparent horizon and $S_B$ is the usual Bekenstein entropy. Also even in modified gravity theories namely in Gauss-Bonnet gravity and in Lovelock gravity, they were able to show the equivalence between the modified Einstein field equations and the thermodynamical laws. As an extension to these works, Cai et al., \cite{Cai:2008,Cai:2005a} studied the unified first law in Einstein gravity as well as in modified gravity theories namely Lovelock gravity, scalar--Tensor theory and Brane--World scenario\cite{Chakraborty:2014,Chakraborty:2014b,Mitra:2015a}. Elling et al.,\cite{Eling:2006a} on the otherhand showed that Clausius relation should be modified by introducing entropy production term in $f(R)$ gravity theory. In the present work we shall modify the horizon entropy suitably so that Clausius relation is automatically satisfied for Rastall gravity theory.\\
The line element for FLRW space time can be written as 
\begin{equation}
ds^2=h_{ab}dx^a dx^b+R^2 d \Omega_2^2, \label{t2}
\end{equation}
or in double--null form as 
\begin{equation}
ds^2=-2dl dm+R^2 d \Omega_2^2, \label{t3}
\end{equation}
where,
\begin{equation}
\partial_{l, m}=\frac{\partial}{\partial_{l, m}}=-\sqrt{2}\left(\partial_t \pm \frac{\sqrt{1-k r^2}}{a}\partial_a\right),\label{t4}
\end{equation}
are future pointing null vectors, $R=ar$ is the geometric radius (or area radius), $h_{ab}=\mbox{diag}(-1, \frac{a^2}{1-k r^2})$ is the metric on the 2--space ($x^0=t, x'=r$) and $k=0, \pm 1$ stands for the curvature scalar.\\
According to Hayward, the trapping horizon (denoted as $R_T$) is defined as $\partial _l R|_{R=R_T}=0$ and hence we have
\begin{equation}
R_T=\frac{1}{\sqrt{H^2+\frac{k}{a^2}}}=R_A.\label{t5}
\end{equation}
For any horizon having area radius $R_h$ the surface gravity is defined as 
\begin{equation}
s_g=\frac{1}{2 \sqrt{-h}}\partial_a \left(\sqrt{-h} h^{ab} \partial_b R\right)|_{R=R_h}\label{t6}
\end{equation}
or explicitly
\begin{equation}
s_g=-\left(\frac{R_h}{R_A}\right)^2\left(\frac{1-\frac{\dot{R}_A}{2HR_A}}{R_h}\right),\label{t7}
\end{equation}
which simplifies for apparent horizon as 
\begin{equation}
s_A=-\frac{(1-\epsilon)}{R_A},\label{t8}
\end{equation}
with $\epsilon=\frac{\dot{R}_A}{2HR_A}.$\\

Note that , for $\epsilon < 1$, the surface gravity $\kappa$ to be negative so that the apparent horizon coincides with the inner trapping horizon (for outer trapping horizon the surface gravity is positive)\cite{Cai:2006}.\\

Now the Misner--sharp energy \cite{Misner:1964}
\begin{equation}
E=\frac{R}{2G}\left(1-h^{ab}\partial_a R \partial_b R\right), \label{t9}
\end{equation}
measures the total energy inside a sphere of radius $R$. although it is a purely geometric quantity but it is related to the structure of the space--time as well as to the Einstein's equations. For FLRW model, Universe bounded by the apparent horizon this energy simplifies to
\begin{equation}
E=\frac{R_A}{2G}=\frac{1}{2G\sqrt{H^2+\frac{k}{a^2}}}.\label{t10}
\end{equation} 
Further, according to Hayward \cite{Hayward:1997} the unified first law
\begin{equation}
dE=A\psi+WdV=A\psi_a dx^a+WdV, \label{t11}
\end{equation}
is equivalent to the $(0, 0)$ component of the (modified) Einstein equations in Rastall gravity. Here $A=4 \pi R^2, V=\frac{4}{3} \pi R^3$ are respectively the surface area  and volume of a sphere of radius $R$, the energy supply vector $\psi_a$ and work density $W$  are formally defined as 
\begin{equation}
\psi_a=T^b_a \partial_b R+W \partial_a R,~W=-\frac{1}{2} T^{ab}h_{ab}.\label{t12}
\end{equation} 

For the present Rastall gravity model we have
\begin{equation}
W=\frac{1}{2}\left(\rho_t-p_t\right)=\frac{1}{2}\left(\rho-p\right)+\frac{1}{2}\left(\rho_e-p_e\right)=W_m+W_e,\label{t13}
\end{equation}
and $\psi=\psi_m+\psi_e$,\\
with
\begin{equation}
\psi_m=-\frac{1}{2}\left(\rho+p\right)HR dt+\frac{1}{2}\left(\rho+p\right)a dr, \label{t14}
\end{equation}
and
\begin{equation}
\psi_e=-\frac{1}{2}\left(\rho_e+p_e\right)HR dt+\frac{1}{2}\left(\rho_e+p_e\right)a dr. \label{t15}
\end{equation}
Now using the double null vectors $\partial_{l, m}$ as the basis vectors in the 2--space, the vector $U$ tangential to the apparent horizon surface can be written as
\begin{equation}
U=U_l \partial_l+U_m \partial_m.\label{t16}
\end{equation}
As by definition trapping horizon is identified as 
$$\partial_l R_T=0,$$
so on the marginal sphere
$$U(\partial_l R_T)=0$$
i.e,$$U_l(\partial_l \partial_l R_T)+U_m(\partial_m \partial_l R_T)=0,$$
or
\begin{equation}
 \frac{U_l}{U_m}=-\frac{\partial_m \partial_l R_T}{\partial_l \partial_l R_T}.\label{t17}
\end{equation}
Considering $R_T=R_A$ for the present model
\begin{equation}
\frac{U_l}{U_m}=\frac{(1-\epsilon)}{\epsilon}. \label{t18}
\end{equation}
Hence in $(r, t)$ co-ordinate $U$ can be written as
\begin{equation}
U=\frac{\partial}{\partial t}-(1-2\epsilon)Hr \frac{\partial}{\partial r}. \label{t19}
\end{equation}

Now, the actual first law of thermodynamics on the apparent horizon can be obtained \cite{Cai:2006} from the above unified first law by projecting along the above tangent vector $U$, i.e.,

\begin{equation}
\big<dE, U\big>=\frac{s_g}{8 \pi G}\big<dA, U\big>+\big<W dV, U\big>.\label{t20}
\end{equation}

As the projection of the pure matter energy supply $A\psi_m$ along the apparent horizon gives the heat flow $\delta Q$ in the Clausius relation $\delta Q=T dS$, so from the above  equation (\ref{t20}) we obtain\cite{Mitra:2015a, Mitra:2015jqa, Mitra:2015nba, Saha:2015gha, Mitra:2015wba, Randall:1999, Randall2:1999, Maartens:2001, Cao:2013}
\begin{equation}
\delta Q=\big<A\psi_m, U\big>=\frac{s_g}{8 \pi G}\big<dA, U\big>-\big<A\psi_e, U\big>,\label{t21}
\end{equation}
For the Rastall gravity, as the effective matter behaves as cosmological parameter so $\rho_e+p_e=0$ i.e., $\psi_e=0$, hence we have 
\begin{equation}
\delta Q=\frac{s_g}{8 \pi G}\big<dA, U\big>.\label{t22}
\end{equation}
Using Hawking temperature on the apparent horizon as 
\begin{equation}
T_A=\frac{|s_A|}{2 \pi}=\frac{1-\epsilon}{2\pi R_A},\label{t23}
\end{equation}
we have 
$$ \delta Q=\big<A \psi_m, U\big>=T_A \frac{d(A_A)}{4G}.$$
Thus Clausius relation gives
$$S_A=\frac{A_A}{4G},$$
which is nothing but the Bekenstein entropy. Hence for Rastall gravity there is no correction  term in the Bekenstein entropy on the apparent horizon. Finally, it should be noted that due to $\psi_{e}=0$ there will no longer be any correction term for the entropy at other horizon also.

\section{ Conclusion }
 
  An explicit cosmological investigation for generalized Rastall gravity is presented in this work. At first it is examined whether any non-singular model of the universe is possible for this gravity theory and it is found that for suitable choice of the Rastall parameter with restriction on the equation of state parameter a model of emergent scenario can be formulated. For homogeneous and isotropic FLRW model this gravity model is shown to be equivalent to Einstein gravity with particle creation mechanism. Interestingly, it is possible to have a complete cosmic scenario from early inflationary era to the present accelerating phase through the matter dominated era of evolution for this model of gravity with a continuous choice of Rastall parameter. Finally universal thermodynamics is formulated with the present gravity theory and there is no modification( or correction) to the Bakenstein entropy not only at the apparent horizon but also at any other horizon.\\
  
  It is well known that for the last few months a debate has been going on the Rastall gravity theory whether it is modified gravity theory or it is essentially an Einstein gravity with some specific choices for the matter content. As a result a lot of curiosity has been developed to study this model and it motivates the authors for the present work. The results of the present study in comparison with Einstein gravity can be stated point wise as follows:\\
  
  $\bullet$ The generalized  Rastall gravity is equivalent to Einstein gravity with non-interacting variable cosmological parameter.\\
 
  $\bullet$ The generalized Rastall gravity is equivalent to Einstein gravity with particle creation mechanism and the particle creation parameter is related to the variable Rastall parameter.\\
  
  $\bullet$ Both generalized Rastall gravity and Einstein gravity with particle creation mechanism \cite{Chakraborty:2014a} exhibit continuous cosmic evolution from inflationary era to the present epoch.\\
  
  $\bullet$ A phase of emergent scenario is possible for generalized Rastall gravity as well as for particle creation mechanism in Einstein gravity \cite{Chakraborty:2014ora}.\\
  
  $\bullet$ In the contest of universal thermodynamics both the gravity theories have same thermodynamical parameters namely Bekenstein entropy and Hawking temperature on the horizon.\\
  
   Therefore, one may conclude that generalized Rastall gravity is equivalent to Einstein gravity in non-equilibrium thermodynamical presentation at least for homogeneous and isotropic FLRW model. 
   
    \section*{Acknowledgments} 
    Author DD thanks Department of Science and Technology
    (DST), Govt. of India for awarding Inspire research fellowship (File No: IF160067). SD thanks to the financial support from Science and Engineering Research Board (SERB),
    Govt. of India, through National Post-Doctoral Fellowship Scheme with
    File No: PDF/2016/001435. Also, SD thankful to the Department
    of Mathematics, Jadavpur University where a part
    of the work was completed. SC thanks Science and Engineering Research Board (SERB) for awarding MATRICS Research Grant support (File No: MTR/2017/000407) and Inter University Center for Astronomy and Astrophysics (IUCAA), Pune, India for their
    warm hospitality as a part of the work was done during a visit.

\end{document}